# Characterization, optical properties and electron(exciton)-phonon interaction in bulk $In_2Se_3$ crystals and InSe nanocrystals in $In_2Se_3$ confinement

Yu.I. Zhirko[1], V.M. Grekhov[1], Z.D. Kovalyuk[2]

[1]*Institute of Physics, National Academy of Sciences of Ukraine, Kiev 03028, Ukraine*
[2]*Chernivtsy Branch of Institute for Material Science Problem NAS of Ukraine, 58001 Chernivtsy, Ukraine*

**Abstract.** Complex electron-microscopic, energy-dispersed and wide-temperature (5–300 K) optical absorption and photoluminescence (PL) investigations are carried out into Bridgeman-grown layered $In_2Se_3$ crystals. It is shown that $In_2Se_3$ crystals as a whole have a homogeneous concentration of In and Se atoms, corresponding with $In_2Se_3$ stoichiometry. Nevertheless, $In_2Se_3$ crystals contain a significant amount of dislocations, on which nano-sized interspersions of "pocket" crystal phases of pure InSe, $In_6Se_7$ and monoclinic red β-Se settle down. Optical wide-temperature investigations of $In_2Se_3$ allow us to do the following: establish the width of the band-gap as $E_g = 1.550$ eV at T = 5 K, the exciton binding energy as $\Delta E_{exc} = 14 \pm 1$ meV; determine the frequency of a half-layer A-phonon ($\nu_1 = 48 \pm 3$ cm$^{-1}$), which takes part in electron (exciton)-phonon interaction; and to evaluate the effective masses of carriers $m^*_c = 0.12\ m_0$, $m^*_v = 0.65\ m_0$ and the dielectric permeability $\varepsilon = 9.9 \pm 0.1$. Finally, "tune" the band-gap and character of the electron (exciton)-phonon interaction of nano-sized 3D InSe crystals confined in an $In_2Se_3$ crystal matrix; influence of an InSe nanocrystal radius (R) and of an ensemble of 3D InSe nanocrystals with different radii for an increase of Γ(T,R) - the exciton emission/absorption half-width line - with temperature and radii of InSe nanocrystals are discussed.

## 1. INTRODUCTION

Due to their physicochemical properties - strong covalent hybrid *sp³*-bonds inside crystal layers and the low van der Waals bonds between them - layered $In_2Se_3$ crystals and MX and ternary MX compounds (where M–In, Ga and X–S, Se) of the A3B6 Periodic Table group are low-dimensional semiconductors. As has been recently shown, MX crystals, and especially InSe crystals, are perspective materials for 2D electronic devices, e.g. field effect transistors [1,2], photodetectors [3-6], phototransistors [7,8], thermoelectric materials [9] and memory cells [10]. In addition, bulk layered MX crystals are good candidates for solid state hydrogen storage [11-13] and supercapacitors [14].

In contrast to the better-known InSe crystals, bulk $In_2Se_3$ crystals have been less often investigated due to technological growth difficulties; they are more inhomogeneous, do not have mirror-like cleavage surfaces and are characterized by different crystal structure modifications.

Parts II and III of this article present the experimental data of our scanning electron-microscopy (SEM), energy-dispersed (EDX) and optical (wide temperature photoluminescence and light absorption) investigations of Bridgeman-grown bulk $In_2Se_3$ crystals. Some aspects of physicochemical bonding in $In_2Se_3$ crystals are discussed in Part IV, together with their optical properties in the vicinity of the band-gap edge.

Part IV also presents a discussion of the determination of a band-gap edge at T = 5 K, the exciton binding energy, the effective mass of carriers, dielectric permeability, the character of electron- and exciton-phonon interactions in bulk $In_2Se_3$ crystals and the role of nano-sized interspersions of InSe phase in the optical properties of $In_2Se_3$ crystals.

Separate effort is devoted in Part IV to the investigation of optical properties (band-gap tuning) and character of exciton-phonon interactions in 3D InSe nanocrystals with radii between 15 and 650 nm in the confinement of an $In_2Se_3$ matrix.

Finally, Part V presents a summary of this work and our conclusions.

## 2. EXPERIMENTAL METHOD

Bulk $In_2Se_3$ crystals as in [15] were grown using the Bridgeman method from a stoichiometric melt, with a temperature gradient of 15 K/cm at the crystallization front and a growth velocity of 1 mm/h. They have n-type conductivity with concentration of carriers n = 2.0·10$^{18}$ cm$^{-3}$.

Characterization of $In_2Se_3$ crystals was carried out using a Vega3 SBU scanning electron microscope (Tescan) equipped with second electron (SE) and back-scattering electron emission (BSE) detectors, and an EDX microanalysis spectrometer (OXFORD Instruments) with an X-act (10 mm$^2$) SDD detector and INCA EDS software.

Measurements of light absorption and photoluminescence (PL) spectra were made using a 0.6 m optical spectrometer MDR-23 (LOMO) with a grating of 600 groves/mm. Investigations of light absorption and PL spectra for T = 5–300 K were made using a helium cryostat A-255 designed at the Institute of Physics National Academy of Sciences of Ukraine (IP NASU). This was equipped with a UTRECS K-43 system, allowing control of the sample temperature within the range 4.2 to 330 K with an accuracy of 0.1 K. Bulk samples of $In_2Se_3$ crystal were prepared for optical investigations and freely mounted on a large finger before being placed into the helium cryostat.

The excitation of PL spectra was performed using a current-wave semiconductor laser with a wavelength of 532 nm (P = 100 mW) equipped by a laser (SL-532-10 Thorlabs) and edge (LP-03-532-RS Semrock's) filters.

A stable power tungsten lamp (P = 60 W) and standard edge filter with transmission region λ > 600 nm were used for light absorption measurements. A liquid nitrogen mini-cryostat system MK-80 (IP NASU) was used to cool a near-infrared photomultiplier tube FEU-62 (400 < λ > 1200 nm), which served as a radiation detector.

The geometry of light absorption near the band gap edge and PL investigations of $In_2Se_3$ crystals are presented in the insets of Figs. 2a) and 2c), respectively.

## 3. RESULTS
### 3.1 SEM and EDX investigations of $In_2Se_3$ crystals

Figs. 1a) and 1c) show SEM images of $In_2Se_3$ crystal surfaces with 5,510× and 90,000× magnification. It can be seen from Fig. 1a that the fresh cleavage surfaces contain spatial dislocations of 100–200 nm in radius, and the interspersion of foreign sub-micron phases of 400–600 nm in size can be observed in the vicinity of these dislocations.

EDX investigations of these interspersions (Fig. 1b at 20,400× magnification) show that they are: i) phases consisting of 100% Se atoms (point 1); ii) phases of InSe (point 3), which contain 50.17% Se atoms and 49.83% In atoms; or iii) phases of $In_6Se_7$ (points 2, 4 and 5). Note that the analytical area for EDX investigation conducted at an electron beam acceleration of 15 κV was 0.53 mkm for In and 1.4 mkm for Se atoms.

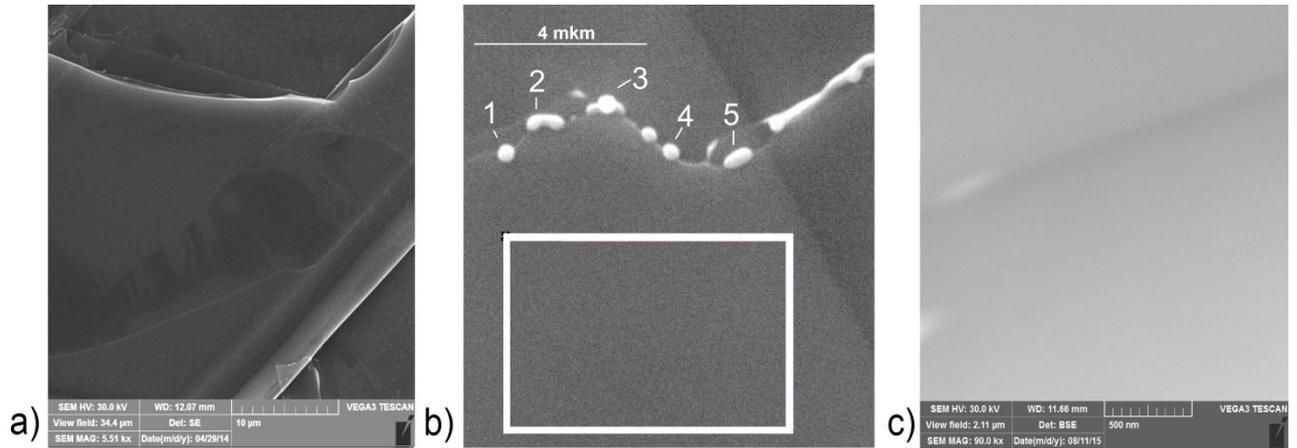

Fig. 1. a) (SE) and c) (BSE): electron-microscopic images of $In_2Se_3$ crystal surfaces with 5,510× and 90,000× magnification; b) EDX investigation of points 1–5 and a selected rectangular area of the same $In_2Se_3$ crystal surface with a 20,400× magnification

With a magnification of 90,000× for the chosen BSE geometry (see Fig. 1c) and a low-current electron beam (the latter was compressed down to 25 nm), one can see dislocation veins of radius 20 to 30 nm and inclusions; these have an ellipsoidal shape with main axes a = b = 30 nm and c = 150 nm.

The cleavage surfaces of the $In_2Se_3$ crystal, which do not contain dislocations, and sub-micron interspersions (see the selected rectangular area in Fig. 1b) may be attributed to pure $In_2Se_3$ crystals.

For the example of GaSe crystals, it has been shown [12] that the interspersion of pure Se phase is a monoclinic modification inherent to the red β-Se (group 2.2/m). Moreover, according to [13], one can observe on the cleavage surfaces of bulk GaSe crystals after natural intercalation in distillated water (Tescan high resolution SEM Lira3 with immersed magnetic lenses) a shrub-like formation of dendrites, consisting of Se atoms grown from the spontaneous nucleation of red β-Se. The spatial size of these shrubs in GaSe crystals after half-year water natural intercalation was about 2–5 mkm, and their whiskers were about 25–50 nm in diameter, tapering to the end with a step of 15 ± 2 nm.

In [16], γ-$In_2Se_3$ nanorods were directly grown on Si (111) substrates. Using high-resolution SEM JSM-7001F (JEOL), it was shown that $In_2Se_3$ nanorods are straight and are not tapered. The average diameter and height of the $In_2Se_3$ nanorods are about 64 and 460 nm, respectively.

Thus, SEM and EDX investigations have shown that Bridgman-grown bulk $In_2Se_3$ crystals contain great numbers of small nanoparticles, which contain different concentrations of In and Se atoms.

### 3.2 Edge absorption and photoluminescence of bulk $In_2Se_3$ crystals

Fig. 2a) shows the light absorption spectra of $In_2Se_3$ crystals at T = 5–300 K with geometry E ⊥ C, k ∥ C (where E is the electromagnetic wave vector and C is the crystal optical axis), with the account of light reflection. In this case, the dielectric permeability of $In_2Se_3$ crystals was chosen as ε = 10.0, which is close to the value of ε = 10.5 for InSe crystals [17]. As the crystal has a number of nano-interspersions, a diameter of 2 mm was chosen for the investigated crystal surfaces and a crystal thickness of 1.2 mm for the light absorption measurement.

We consider two independent series of absorption measurements. In the first case, the width of the spectrometer entrance slit was 0.1 mm, and in the second it was 0.8 mm. This allowed us to carry out practical absorption measurements of the $In_2Se_3$ crystal matrix without accounting for the absorption of light by foreign interspersions. For these cases, the spectral slit width during the experiment did not exceed 0.33 and 2.7 meV. We obtain identical absorption spectra in both cases.

At T = 5 K, light absorption in the bulk $In_2Se_3$ crystal rapidly increases at a wavelength of λ < 800 nm. At λ > 800 nm, a tail of electron density states of the conductive band was observed. This allows us to determine the band gap of $In_2Se_3$ crystal with sufficient accuracy at T = 5 K, at a value of $E_g$ = 1.550 ± 0.001 eV. This value is in good agreement with that in [18], where $E_g$ = 1.56 eV was calculated for an α-$In_2Se_3$ crystal at T = 4.2 K.

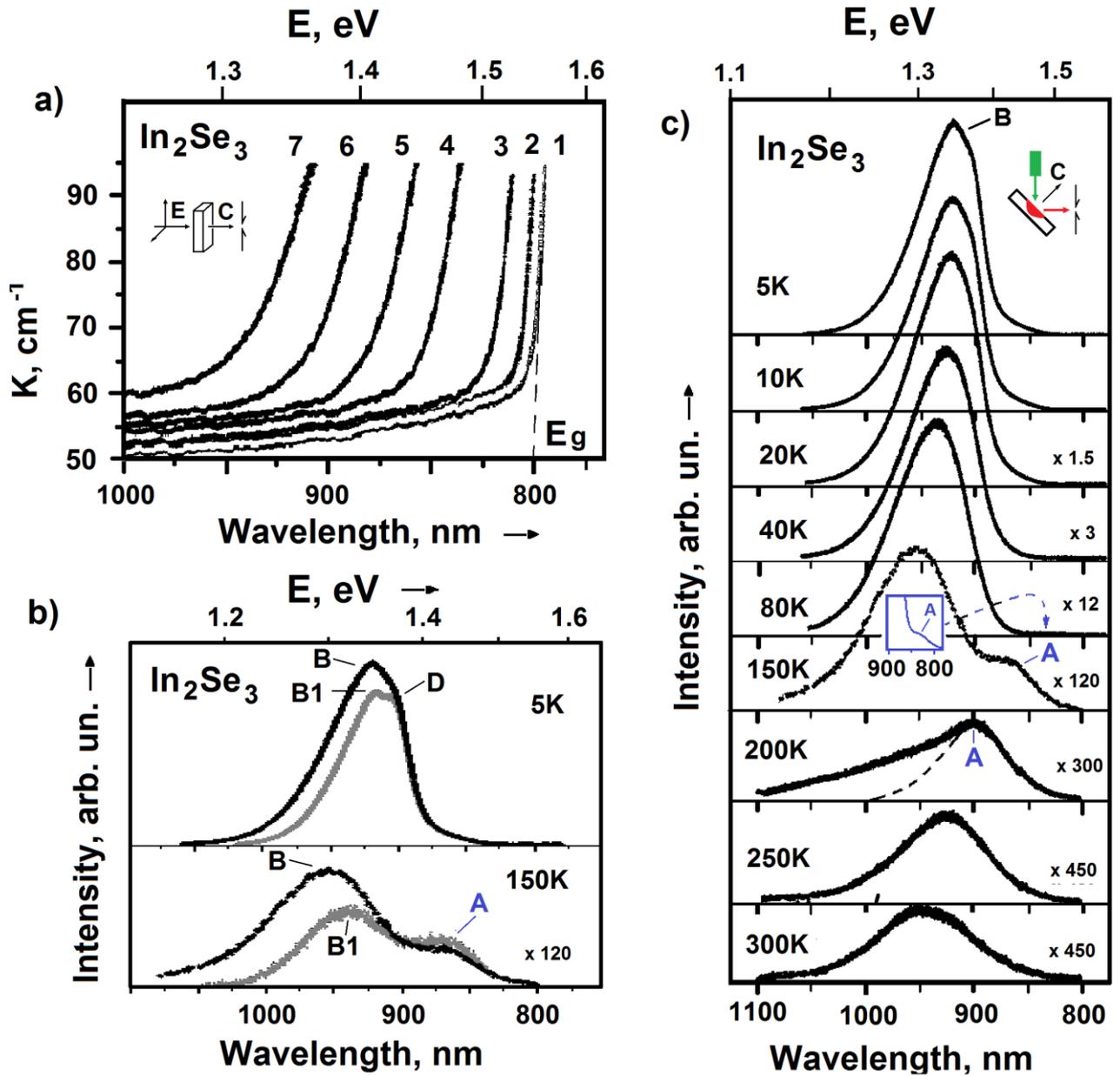

Fig. 2. a) Fundamental band-gap edge of bulk In$_2$Se$_3$ crystals light absorption at: 1) 5 K, 2) 40 K, 3) 80 K, 4) 150 K, 5) 200 K, 6) 250 K and 7) 300 K; b) PL spectra for two different points on the In$_2$Se$_3$ crystal at T = 5 K and at T = 150 K with intensity magnification ×120; c) PL spectra of In$_2$Se$_3$ crystals at T = 5–300 K, shown with intensity magnification between × 1.5 at T = 20 K and × 450 at 300 K.

Moreover, in [19], PL investigations show that pure α-In$_2$Se$_3$ crystals exhibit two PL bands at liquid helium temperature: one at a higher energy of 1.523 eV, due to the radiative recombination of impurity-bound excitons, and the other at a lower energy of 1.326 eV, due to recombination on luminescence centers formed by intrinsic defects due to disorder in the cation sublattice.

A temperature increase (Fig. 2a) of up to 300 K leads to a long-wave shift in absorption edge, accompanied by a reduction in the absorption coefficient inclination corner caused by electron temperature redistribution ($k_BT$) in the tail of the conductive band (CB). Furthermore, with an increase in temperature, an increase of light scattering on crystal lattice vibrations was observed in the forbidden gap.

PL spectra were investigated from different points in the same In$_2$Se$_3$ crystal. The diameter of the laser beam on the crystal surface was 1.5 mm, and the size of the spectrometer entrance slit up to 80 K did not exceed 50 mkm with a spectral resolution of 0.17 meV.

Fig. 2b) shows the PL spectra of bulk In$_2$Se$_3$ crystal for two different points, at T = 5 and 150 K. At T = 5 K, the PL spectra for the first point (black curve) consist of a wide band named the B band, with a peak at 1.3476 eV, and for the second point (grey curve) a B1 band with a peak at 1.3511 eV. In addition, on the high-energy side of the B1 band, a shoulder D at 1.3666 eV was observed.

Fig. 2c) shows the PL spectra from the first point on the crystal surface of the bulk In$_2$Se$_3$ crystal at T = 5–300 K. It can be seen that at T

> 200 K, the B band practically disappears and a new high-energy A band starts to prevail, which begins to be detected at T = 80 K (see insert in Fig. 2c).

The temperature shift of the band-gap width ($E_g$) and the energetic position maxima for the A, B, and B1 bands are presented in Table 1.

TABLE 1. Energy positions of the B, B1 and A bands, the band-gap $E_g$, and a half-width on a half maximum Γ of the B and B1 bands in the bulk $In_2Se_3$ crystal versus temperature

| Temperature, K | B, eV | B1, eV | A, eV | $E_g$, eV | $Γ^B$, meV | $Γ^{B1}$, meV |
|---|---|---|---|---|---|---|
| 5 | 1.3476 | 1.3511 | - | 1.5490 | 45.0 | 40.5 |
| 10 | 1.3467 | 1.3525 | - | 1.5474 | 45.5 | 41.2 |
| 15 | - | 1.3505 | - | - | - | 41.9 |
| 20 | 1.3447 | - | - | 1.5430 | 46.5 | - |
| 25 | - | 1.3486 | - | - | - | 42.7 |
| 40 | 1.3382 | 1.3469 | - | 1.5331 | 49.0 | 44.5 |
| 70 | - | 1.3389 | - | - | - | 47.1 |
| 80 | 1.3264 | - | 1.4937 | 1.5070 | 54.5 | - |
| 100 | - | 1.3295 | - | - | - | 48.3 |
| 150 | 1.2982 | 1.3202 | 1.4301 | 1.4450 | 67.0 | 47.8 |
| 200 | - | - | 1.3930 | 1.4057 | - | - |
| 250 | - | - | 1.3458 | 1.3620 | - | - |
| 300 | - | - | 1.3131 | 1.3274 | - | - |

## 4. DISCUSSION

In Fig. 3, the open squares indicate the experimental dependence of $E_g(T)$, obtained from absorption measurements of bulk $In_2Se_3$ crystals. The open circles and triangles show the energetic shift in the PL spectra of the A and B bands versus temperature. It can be seen that the temperature shift of $E_g$ and the A band are described by identical curves, 1 and 2, which are separated by 14 ± 1 meV. At the same time, the temperature shift of the B band essentially differs from the temperature shift of $E_g$ and the A band.

It is known [18, 20] that single $In_2Se_3$ crystals grown by the Bridgman method possess four modifications (α, β, γ, and δ), which differ from each other in terms of their sequence of crystalline layer stacking. Each crystal layer consists of five monolayers of Se – In – Se – In – Se atomic sheets, bounded by hybrid $sp^3$ covalent bonds. Interatomic distances between atomic sheets within a layer are as follows: i) In–Se: 0.287 nm (In situated in octahedral holes at 523 K) and 0.269 nm (In situated in tetrahedral holes at room temperature); and ii) Se–Se: 0.401 nm (Se between each layer) [18].

The space group of the β modification are rhombohedral $D^5_{3d}$. The unit cell has dimensions a = 0.405 nm, C = 2.941 nm at 523 K, and Z = 3. The unit cell of the α modification ($C^5_{3v}$ – space group) has dimensions a = 0.405 nm, C = 2.877 nm at room temperature, and Z = 3.

For bulk GaSe and InSe single crystals, the variation in $E_g(T)$ is mainly attributed to a change in the electron-phonon interaction (the self-energy effect) due to the participation of a homopolar half-layer $A^1_{1g}$-phonon, which reduces the band-gap width when the temperature increases, as shown in [21-23]. In this case, the self-energy contribution obeys the equation

$$\Delta E_g(T) = \frac{8ln2}{\pi}\left[\frac{g_v^2}{\sqrt{2m_v^*}} + \frac{g_c^2}{\sqrt{2m_c^*}}\right]\hbar n^* Q\sqrt{\hbar\omega}, \quad (1)$$

where the subscripts $v$ and $c$ refer to the valence band (VB) and conductive band (CB), $n^*$ is the number of homopolar phonons in that particular mode, $g_v$ and $g_c$ are the coupling constants, $m^*_v$ and $m^*_c$ are the effective masses of carriers, and $Q$ is the radius of the Brillouin zone. The energy of the homopolar phonon is taken to be constant over the whole Brillouin zone.

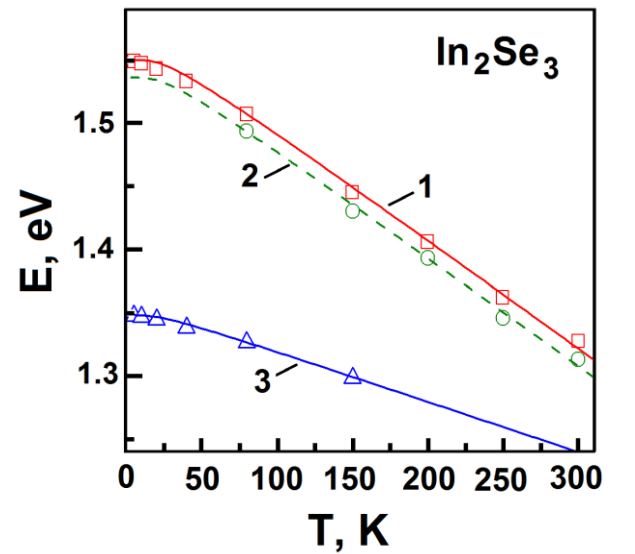

Fig. 3. Open squares, circles and triangles show experimental data for $E_g$, A, and B band shifts versus temperature. Curves 1 and 2 correspond to Eqs. (3) and (3′), and curve 3 corresponds to Eq. (4).

According to [21], using crystal parameters of $a = 0.4001$ nm and $c = 2.5232$ nm for γ-InSe crystals ($C_{3v}^5$ space group) [24], values of effective mass for carriers $m^*_c = 0.135\ m_0$ and $m^*_v = 0.7\ m_0$ in CB and VB, and exciton binding energy $\Delta E_{exc} = 14.5$ meV for InSe [17,25] an analytical expression for $E_g(T)$ dependence can be obtained for bulk InSe crystals due to the self-energy effect with homopolar $A^1_{1g}$ phonon ($\hbar\omega = 13.6$ meV [26]) participation as follows:

$$E_g^{InSe}(T)[meV] = 1351.2 - \frac{65}{\exp\left(\frac{162}{T}\right)-1} \quad (2)$$

Note that in [21], the temperature dependency of the $E_{n=1}$–ground exciton state shift was investigated for the case of exciton-phonon interaction, and an averages exciton-phonon coupling constant $g = 0.5$ (for electrons in CB and holes in VB) was found.

Following [21,17], and taking into account the crystal parameters of bulk $In_2Se_3$ crystal [18] and our experimental data for $E_g(T)$ dependence (see Table 1), an analytical dependence

$$E_g^{In_2Se_3}(T)[meV] = 1550 - \frac{60}{\exp\left(\frac{70}{T}\right)-1} \quad (3)$$

was found for bulk $In_2Se_3$ crystals with $g = 0.65$ ($g_c = g_v = 0.65$), $\hbar\omega = 6.0$ meV and $m^*_c = 0.12\ m_0$, $m^*_v = 0.6\ m_0$. Note that the same analytical dependence

$$E_A(T)[meV] = E_g^{In_2Se_3}(T)[meV] - 14 \quad (3')$$

was found for the A-band shift versus temperature.

The experimental data for $E_g(T)$ and $A(T)$ dependences are in accordance with Eqs. (3) and (3'), and differ one from another at a constant value of 14 meV. This allows us to make the assumption that the A band displays free exciton emission in bulk $In_2Se_3$ crystals, with an exciton binding energy $\Delta E_{exc} = 14 \pm 1$ meV. Note that the obtained value is very close to the exciton binding energy $\Delta E_{exc} = 14.5$ meV in bulk InSe crystals.

Now, using the obtained fitting values for $m^*_c$ and $m^*_v$ and the well-known equation for exciton binding energy $\Delta E_{exc}[meV] = 13605 \cdot \mu^*/\varepsilon^2$, where $\mu^* = (m^*_c \cdot m^*_v)/(m^*_c + m^*_v)$ is the exciton effective mass, a value for the dielectric permeability of $\varepsilon = 9.9 \pm 0.1$ in bulk $In_2Se_3$ crystals can be obtained. This value is also close to $\varepsilon = 10.5$ [17] in bulk InSe crystal.

It can be seen that in the case of $In_2Se_3$ crystals, the temperature shift of $E_g$ occurs with the participation of phonons whose energy ($\hbar\omega = 6.0$ meV) and vibration frequency ($\nu_1 = 48$ cm$^{-1}$) are significantly lower than the energy ($\hbar\omega = 13.6$ meV) and the frequency ($\nu = 110$ cm$^{-1}$) of a half-layer homopolar $A^1_{1g}$ - phonon in the bulk InSe crystal.

This reduction in frequency vibration for a half-layer homopolar phonon in an $In_2Se_3$ crystal is caused by:

i) An increase in the atomic mass of the $In_2Se_3$ molecule, as compared to the $In_2Se_2$ molecule of InSe crystal. This increase in atomic mass in a simple model of one-dimensional linear chain corresponds with the equation $\nu_1 = \nu/[1 - (M - M1)/M]^{1/2}$ (where $\nu_1$, M1 are the frequency and atomic mass of the $In_2Se_3$ molecule, and ν, M are those of $In_2Se_2$) reducing the frequency for the half-layer homopolar phonon in $In_2Se_3$ crystal to $\nu_1 = 96$ cm$^{-1}$;

ii) The difference in the distribution of valence electrons between the In and Se atoms in molecules of InSe and $In_2Se_3$ crystals. As can be seen from Fig. 4, the transition from the InSe crystal to the $In_2Se_3$ crystal is accompanied by the appearance of an additional Se atom between two intra-layer In atoms, which takes up missing electrons from each of its six neighboring In atoms. This internal covalent $sp^3$ bond between In→Ga←In atoms (six nearest-neighbor In atoms) in a layer of $In_2Se_3$ crystal is much weaker than the (2s) double electron In<=>In covalent bond in the InSe crystal. This leads, in accordance with the Hooks law, to reduction of the half-layer homopolar vibration frequency and increase layer thickness.

Note that according to [27], optical vibration at $\nu = 54$ cm$^{-1}$ was observed in the Raman spectra of γ-$In_2Se_3$ crystals. In [28], a weak singularity at 41 cm$^{-1}$ with damping parameter γ = 97 cm$^{-1}$ for α-$In_2Se_3$ crystals was attributed to a plasmon contribution to the far infrared spectrum.

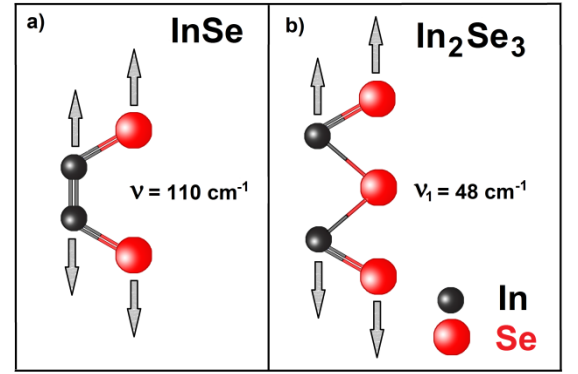

Fig.4. Sketch of homopolar half-layer vibrations and covalent bonding in InSe and $In_2Se_3$ crystals

Based on the experimental data for $E_g(T)$ dependency, and the calculations and discussion above, it is reasonable to suggest that the frequency of a half-layer homopolar A-phonon in Bridgeman-grown bulk $In_2Se_3$ crystal is $\nu_1 = 48 \pm 3$ cm$^{-1}$.

**4.1 B band emission**
As can be seen in Fig. 2c), the PL intensity of the B band prevails at T=5 K and decreases with temperature, while the PL intensity of the A band at T > 80 K increases with temperature. At T=5 K, the B band for some crystal points has a structure consisting of a B1 band and a shoulder D, whose energetic position slightly differs from the B band (see Fig. 2b).

In Fig. 3, the energy shift of the B band versus temperature is shown by open triangles. It can be seen that the energy position of the B band at T = 5 K ($E_B = 1.3476$ eV) and the energy shift of the B band with temperature significantly differ from the energy position and shift of the A band and $E_g$.

The energy shift with temperature for the B band is well approximated by curve 3, based on the following analytical dependence:

$$E_B(T)[meV] = 1347.6 - \frac{24}{\exp(\frac{60}{T})-1} \quad . \quad (4)$$

Curve 3 satisfactorily describes the temperature shift of the B band with participation of phonons possessing energy $\hbar\omega = 5.1$ meV ($\nu = 41$ cm$^{-1}$). This half-layer transversal vibration of E-symmetry with frequency $\nu = 40$ cm$^{-1}$ takes place in InSe crystals and is pronounced in IR and Raman spectra [29,30].

In the following discussion, we assume that the B band:

i) Is caused by radiative recombination of excitons in InSe nanocrystals;

ii) Has an emission peak at T = 5 K which is shifted by 11 meV to the short-wave side relative to the emission line of the ground exciton state ($E_{n=1}$ = 1336.7 meV) in bulk crystals γ-InSe [31]; and

iii) Has a temperature shift which occurs with the participation of polar half-layer phonons of E-symmetry, which has a frequency $\nu = 40$ cm$^{-1}$ similar to that in bulk crystals, in the electron-phonon interaction process.

In this case, the half-layer E-vibration in InSe nanocrystals with frequency $\nu = 40$ cm$^{-1}$ is almost in resonance with the fully symmetric half-layer A-vibration of the In$_2$Se$_3$ matrix of $\nu_1 = 48$ cm$^{-1}$ (Fig. 4).

It was shown in [31] that the availability of layer stacking faults (SF) - a mixture of γ- and ε-modifications - results in the appearance of additional bands at T = 4.5 K in the spectra of exciton photoluminescence which are 1–1.5 meV higher than the energy of the ground exciton state $E_{n=1}$ = 1336.7 meV in γ-InSe single crystals. In ε-GaSe crystals, the availability of SF leads to the appearance of a thin structure related to free excitons localized at SF and which is below the energy of the ground exciton state in ε-GaSe single crystals [32-34].

However, the increase in free exciton recombination energy of 11 meV observed in InSe nanocrystals is problematic to explain based on the availability of stacking faults, since for the InSe nanocrystal sizes shown in Fig. 1c) these stacking faults should, as a rule, extend from the nanocrystal bulk to its surface.

A more likely reason for the high-energy shift of the exciton band is therefore the so-called "blue" shift, caused by a decrease in the geometrical sizes of bulk crystals to those of nanocrystalline particles.

Based on this assumption, we perform some calculations. For γ-InSe, the volume of a unit cell to a good approximation is $V_0 \approx 0.4$ nm$^3$, which corresponds to the volume of a sphere with radius $r_s$ = 0.457 nm. Then, from the inequality defining the ratio of a nanocrystal surface's to its volume

$$N^3/2 > 6N^2, \quad (5)$$

where $N$ is the number of unit cells along one of the faces for a nanocrystal of the cubic shape, it follows that Eq. (5) is valid for an InSe spherical particle with radius $R^{QD} > 5.5$ nm. Note that the Rydberg radius of a Wannier-Mott exciton in InSe bulk crystals is $r_{exc}$ = 4.5 nm. Hence, for shown in Fig. 1c) (SEM images) InSe nanocrystals with radius $R^{QD}$ = 30 nm, the number of unit cells is $(R^{QD}/r_s)^3 \approx 2.8 \times 10^5$.

For the bulk InSe crystals, the average values of the anisotropic effective masses of electrons and holes are $m_e$ = 0.14$m_0$ and $m_h$ = 0.72$m_0$ [35,36], while their Rydberg radii are $r_e$ = 4.0 nm and $r_h$ = 0.8 nm, respectively.

We then suppose that the following condition is valid for InSe nanocrystals at $R^{QD} > 5.5$ nm:

$$R^{QD} > r_{exc} > r_e > r_h \quad , \quad (6)$$

which allows a description of phonon branches, exciton behavior and exciton binding energy as in a classic case of bulk InSe crystals, based on quantum-dimensional corrections in the first approximation, while the behavior of charge carriers should be considered within the framework of quasi-3D conduction and valent minibands.

In the case where $R^{QD} < r_{exc}$, radiative recombination will be observed of an electron-hole pair non-bound by Coulomb interaction. Here, it will be important to take into account the ratio of the radii of charge carriers and the quantum dot.

For the set $i$ = 1, ... $n$ of nanoparticles possessing various radii $R_i^{QD}$, the following relationship is valid:

$$E_i^{QD}[eV] = \sqrt{E_g^2 + \frac{2\hbar^2 E_g \pi^2}{\mu (R_i^{QD})^2}} \quad , \quad (7)$$

that is obtained using the hyperbolic model for the blueshift of the forbidden gap inherent to the crystal when its geometrical dimensions are decreased [37], where $E_i^{QD}$ is the forbidden gap width corresponding to the nanocrystal of the radius $R_i^{QD}$.

In accordance with Eq. (7), Figs. 5a) and 5b) show the fitting of calculated and experimental data obtained at T = 5 K from the measured PL spectra of the B and B1 bands.

The fitted spectra well approximate the B and B1 bands by the set of six Lorentz lines L1–L6 of different intensities (which describe the emission of Wannier-Mott free excitons in InSe nanocrystals with radii within the range 16.2 to 627 nm) and Gauss lines G1, G2 (describing the emission of excitons bound by defects at the boundary between InSe nanocrystals and the crystalline matrix In$_2$Se$_3$). It should be noted that in the case of the B1 band, the weakly intensive G2 line is practically absent.

Table 2 gives the calculated values of the fitted half-widths at half-heights $\Gamma_{Li,Gi}$, the energy positions of peaks $E_{Li,Gi}$ and the integrated intensities $S_{Li,Gi}$ of the L and G lines of emission inherent to free and bound excitons, respectively, for InSe nanocrystals of various radii at T = 5 K.

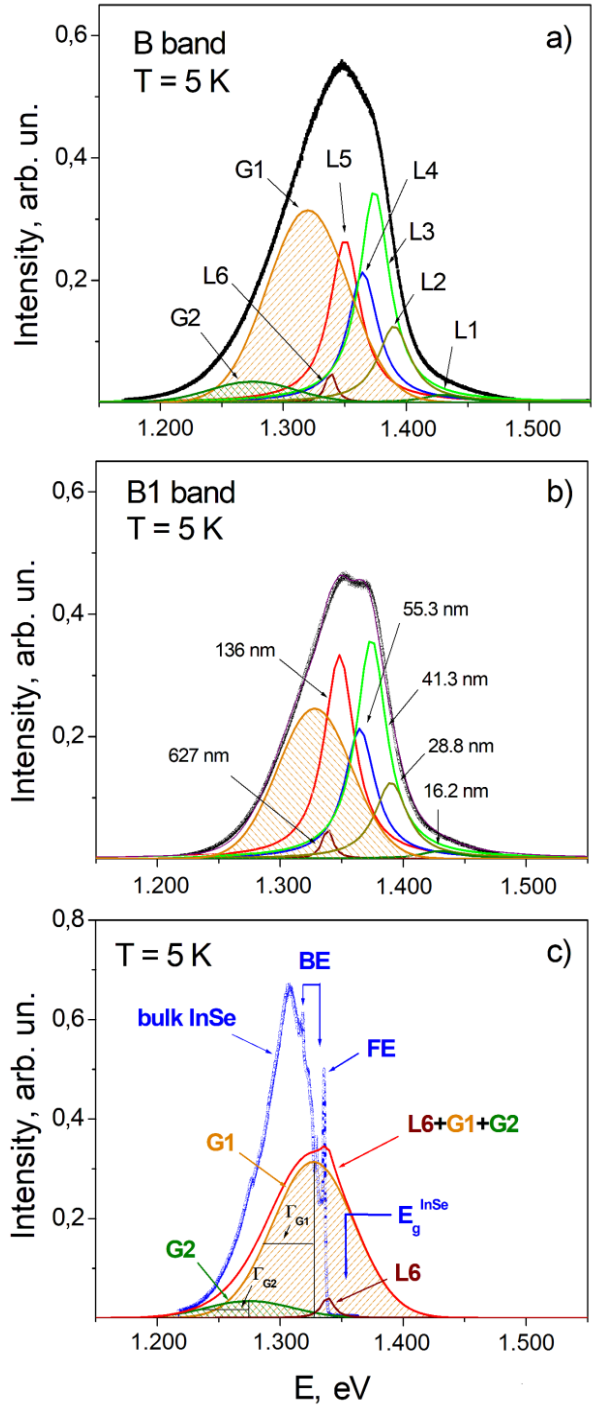

centers, formed by intrinsic defects caused by disorder in the cation sublattice of $In_2Se_3$.

In these calculations, we assume that, even for small sizes of InSe nanoparticles, the behavior of electrons and holes inside energy minibands and their Coulomb interaction can be described as for the case of bulk crystals. The radii of InSe nanocrystals (Table 2) were therefore calculated using Eq. (7) and the expression

$$E_i^{QD}[eV] = E_{Li} + \Delta E_{exc}^{InSe}, \qquad (8)$$

where $\Delta E_{exc}^{InSe} = 0.0145$ eV is the binding energy of excitons in bulk InSe single crystals.

At the same time, it should be noted that in the layered InSe and GaSe crystals, intercalated with hydrogen at concentrations x > 1.0 for InSe [12] and x > 2.0 for GaSe [38], non-linear behavior can be observed in value of the exciton binding energy, which is related to the filling of the van der Waals gap by hydrogen molecules and the increase in dielectric permittivity in the interlayer space.

Table 2. Calculated values of half-widths at half-heights ($\Gamma$), integrated intensities ($S$) and peak positions ($E$) for emission lines of free and bound excitons in InSe nanocrystals of radii $R_i^{QD}{}_{InSe}$, obtained by approximation of B and B1 emission bands at T = 5 K

| Line number | $R_i^{QD}{}_{InSe}$ | $E_{Li,Gj}$ | $\Gamma_{Li}, \Gamma_{Gj}$ | B1 band $S_{Li,Gj}$ | B band $S_{Li,Gj}$ |
|---|---|---|---|---|---|
| | [nm] | [eV] | [meV] | [arb. un.] | |
| Lorentzian shape lines of free excitons in InSe nanocrystals | | | | | |
| L1 | 16.2 | 1.4300 | 20 | 5.0 | 5.0 |
| L2 | 28.8 | 1.3900 | 16 | 32.0 | 32,0 |
| L3 | 41.3 | 1.3740 | 15 | 81.0 | 78.0 |
| L4 | 55.3 | 1.3646 | 15 | 48.0 | 48.0 |
| L5 | 136.0 | 1.3480 | 15 | 75.0 | 71.0 |
| L6 | 627.0 | 1.3385 | 7 | 1.75 | 1.75 |
| Gaussian shape lines of bond excitons in InSe nanocrystals | | | | | |
| G1(B1) | | 1.3270 | 30 | 525 | |
| G1(B) | | 1.3270 | 33 | | 858 |
| G2(B) | | 1.2750 | 35 | | 105 |

Fig. 5. PL spectra of an $In_2Se_3$ crystal at T = 5 K: a), b) B and B1 bands and emission $L_i$ lines of free excitons in InSe nanocrystals of radii $R_i^{QD}$, calculated based on Eqs (7) to (10), and $G_i$ emission lines of excitons bound with defects at the boundary between the InSe nanocrystal and $In_2Se_3$ matrix; c) PL spectrum of the InSe single crystal (where FE, BE are emission lines of free and bound excitons in bulk InSe single crystals, respectively) and calculated L6, G1 and G2 lines for InSe nanocrystal. $E_g^{InSe}$ is the band gap energy of a bulk InSe crystal at T = 5 K

For comparison, Fig. 5c shows the PL spectrum (T = 5 K) of a specially non-doped InSe single crystal and the fitting spectra of the Lorentz L6 and Gauss G1, G2 components of the emission B band observed in InSe nanocrystals.

It should be noted that the fitted G1 line is close in energy to the PL band observed in [19] at 1.326 eV; this is related to recombination on luminescence

In accordance with [12,38], this leads from the very beginning to an increase in the exciton radius and, with the subsequent increase in hydrogen molecules concentration in the interlayer space, to the localization of exciton motion within the planes of the crystalline layers. Since in InSe, excitons comprise 12 crystalline layers and in GaSe only seven layers, the 2D localization of exciton motion in GaSe crystals can be observed only at higher concentrations of hydrogen.

It should also be noted that in studies of the optical properties of InSe nanoflakes of thickness d = 0.8–4.0 nm, which comprises one to five crystalline layers, beside the observed tuning of the forbidden gap [39-41], Coulomb interaction is available between 2D localized electrons and holes in these layers with the creation of 2D excitons. This should also result in an increase in the binding energy of 2D excitons.

In this approximation, it can be seen from the fitting spectra for the B and B1 bands at T = 5 K, shown in Fig. 5, and from the parameters given in Table 2, that:

- The values of Γ for the Lorentz L lines are correlated and increase, with a decrease in the nanoparticle radius, from Γ = 15 meV for $R^{QD} < 150$ nm, to Γ = 16–20 meV for $R^{QD} < 30$ nm;
- The obtained values for InSe nanoparticle radii correlate with the experimental values for the observed nano-crystalline InSe inclusions in the $In_2Se_3$ matrix, as shown in Fig. 1c);
- For $R^{QD} < 20$ nm, the intensity of emission lines due to inclusions of InSe nanocrystals in the matrix of $In_2Se_3$ (at T = 5 K) drops sharply; for $R^{QD} \leq 7$ nm, this intensity is equal to zero, in accordance with Eq. (7), since the energy of fundamental transition in the InSe nanocrystal $E_g$ = 1.564 eV exceeds the forbidden gap width in the $In_2Se_3$ crystal.

Fig. 6 shows the dependences of the integrated intensities $S_{Li}$ for the B and B1 bands on the radius $R_i^{QD}$ of the InSe nanocrystal at T = 5 K. It can be seen that the highest contribution to the emission of B and B1 bands is provided by free excitons in InSe nanocrystals of radii 28.8 to 136 nm; this correlates well with the results of SEM investigations. The insert in Fig. 6 shows the dependence of $S_{Li}$ on $R_i^{QD}$ for InSe nanocrystals in the $In_2Se_3$ matrix; due to their sizes, this corresponds to the range 3D QD.

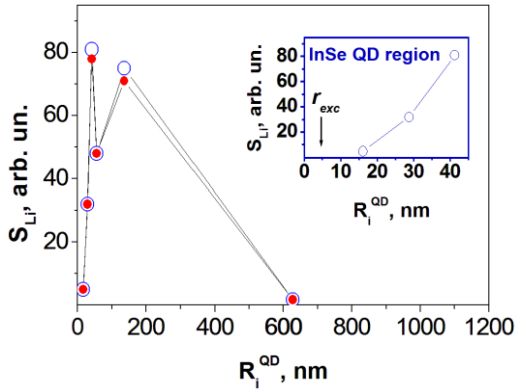

Fig. 6. Dependence of the integrated intensity $S_{Li}$ of the emission lines of free excitons on the radius of the InSe nanocrystal. The arrow in the insert indicates the exciton radius in the bulk InSe crystal

The above discussion allows us to determine the value of the radius cutoff in InSe nanocrystals, at which the emission of free excitons in the $In_2Se_3$ matrix takes place, as $R^{QD}_{min} \approx 15 \pm 1$ nm. The maximum value of the forbidden miniband width in these nanocrystals, at which radiative recombination of free excitons in InSe nanocrystals embedded into the $In_2Se_3$ matrix is observed, can be determined as $E^{QD}_{max} < 1.450 \pm 0.001$ eV.

**4.2 Scheme of energy positions for minibands in InSe nanocrystals embedded into bulk $In_2Se_3$ crystals**

Figs 7a) and 7b) illustrate the scheme of energy positions for valence (MVB) and conductive (MCB) minibands in InSe nanocrystals relative to the valence (VB) and conductive bands (CB) of the $In_2Se_3$ matrix at T = 5 K and T = 300 K. The scheme is plotted in accordance with the analytical dependences obtained in this work for the energy shifts of the forbidden gap $E_g$ and the A, B, B1 PL bands on temperature, and for the redistribution of intensities between the A, B and B1 bands, observed experimentally at T > 80 K, with an increase in temperature of the $In_2Se_3$ crystal.

We assume here that in bulk $In_2Se_3$ crystals, in a similar way to InSe crystals, the main contribution to the change in the forbidden gap width with an increase in temperature is provided by the conduction band. It therefore follows that at T = 5 K in InSe nanocrystals of large radius (> 627 nm), the MCB minimum is localized near $100 \pm 1$ meV, below the conduction band of bulk $In_2Se_3$; the MVB maximum for InSe nanocrystals is $99 \pm 1$ meV higher than the maximum in the valence band in bulk $In_2Se_3$. Presented in the same place is the scheme of the energy positions for VB and CB of bulk InSe, the extremes of which practically coincide in terms of energy with those of InSe nanocrystals of large radii (> 627 nm).

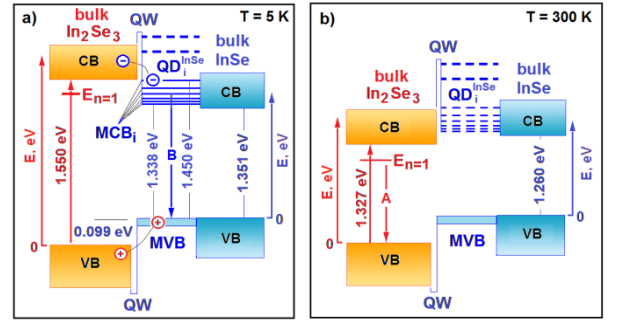

Fig. 7. Scheme of energy positions for MCV and MVB of InSe nanocrystals of various radii relative to VB and CB of bulk crystals $In_2Se_3$ and InSe at temperatures of a) 5 K; b) 300 K

As can be seen in Fig. 7a), both CB electrons and VB holes in $In_2Se_3$ matrix can sufficiently quickly, under optical excitation of electrons into the conduction band, overcome the quantum barriers separating them from InSe nanocrystal minibands located at lower energies. Therefore, at T = 5 K, recombination of hot carriers in $In_2Se_3$ matrix is realized via InSe nanocrystal minibands with the emission of free and bound excitons. The reverse process of carrier transition from the InSe nanocrystal minibands into the VB and CB of the $In_2Se_3$ matrix is inhibited due to cooling of the carriers and the availability of quantum walls (QW) between the MCV and CV, and MVB and VB, respectively.

When the temperature increases to above T = 150 to 200 K, the CB of $In_2Se_3$ crystal is reduced to below the $MCB_i$ of InSe nanocrystal of the radius $R_i^{QD}$. This results in closing the channel for radiative recombination of carriers via $MCB_i$ of InSe nanocrystal of the radius $R_i^{QD}$, and, as shown in Fig. 7b), the wide A band inherent to emission of free excitons in $In_2Se_3$ matrix can be detected in the PL spectra at T > 80 K.

It is worth noting that the QW shown in Fig. 7 are, in the simplest case, a set of defects in the crystalline lattice of InSe nanocrystals and the $In_2Se_3$ matrix.

These regions provide binding of free excitons, which can be seen as the G1 and G2 emission lines in Fig. 5.

The size of the QW is commensurate with the thickness of several crystalline layers (1–2 nm). The simplest defects in bulk InSe and GaSe, responsible for the creation of these QW (donors and acceptors), are shown in Fig. 8.

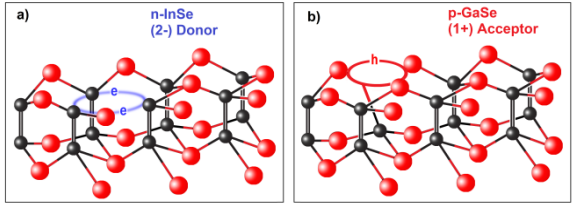

Fig. 8. Simplest donors and acceptors in layered crystals of group $A^3B^6$

Note that in [42] it was reported on the basis of SEM, EDX and EBSD investigations that Bridgman-grown bulk GaSe single crystals contain numerous inclusions of residual monoclinic β-Se (Laue group 2.2/m) in the interlayer space. At the same time, the average proportions of Ga atoms (53%) and Se (47%) essentially differ from the stoichiometric composition. In addition, GaSe crystals possess ideally mirrored split surfaces, are high-Ohmic, and the concentration of free carriers at room temperature lies within the range $p = 10^{13}$ to $10^{14}$ cm$^{-3}$.

It therefore follows that the presence of only point defects is obviously insufficient to describe both the disturbances in stoichiometric composition and the existence of a rather perfect crystalline structure. In view of these conditions, and considering the evolution of the simplest defect of the acceptor type, calculations were performed in [42] for the types of defects that can disturb crystal stoichiometry and cause the aggregation of simple point defects into complex plane and bulk spatial defects during the process of synthesizing the GaSe crystal.

Following [42], the layered crystals of the A3B6 group ordering the atoms both inside crystal layers and between them have the appearance …[X-M-M-X]…WB…[X-M-M-X]…, where WB is the van der Waals bond between [X-M-M-X] crystal layers. Therefore, the loss of one of the metal atoms (M) in the course of synthesis can cause splitting of the [X-M-M-X] layer and give rise to layer bifurcation (or creation of two layers) or, vice versa, the "agglutination" of two layers into one. These areas are characterized by the creation of residual Se atoms "pocket" phases and a considerable amount of Frenkel and Schottky defects, both point and extended, along the perimeter of splitting/agglutination. Thus, in GaSe crystals of the *p*-type, agglutination takes place when three Ga and seven Se atoms are lost in every adjacent crystalline layer in the course of growing.

High concentrations of agglutinations lead to disturbances in stoichiometric composition as a whole; however, in the remaining areas, the crystal remains perfect from a crystallographic viewpoint, since deformations decay quickly owing to the weakness of van der Waals bonds [42].

This splicing/splitting of layers also plays an essential role in the synthesis process of n-In$_2$Se$_3$ crystals and, as seen in Fig. 1, results in fallout of the nanocrystalline "pocket" phases of InSe, In$_6$Se$_7$ and red β-Se between the In$_2$Se$_3$ layers. In addition, it leads to energy band bending and the appearance of local and gap vibrations in the crystalline lattice.

**4.3 Exciton-phonon interaction in InSe 3D quantum dots**

Exciton-phonon interaction in InSe and GaSe crystals has been sufficiently well studied, both within the framework of the classical model by Elliott-Toyozawa [21,43,44] and based on polariton mechanisms of energy dissipation (see [17]).

Since the dimensions of nanocrystals are five to seven orders of magnitude smaller than those of bulk crystals, the manifestation of polariton effects in the absorption/emission of free excitons is usually neglected (with account of essential decay of excitons), and the classical model is used. According to [21], in the case of weak exciton-phonon interaction (Wannier-Mott excitons), the half-width Γ of the absorption line for the ground $n = 1$ exciton state with an increase of crystal temperature is described by the expression

$$\Gamma(T) = \sqrt{[\Gamma_0^h(1+\beta n^*(T))]^2 + (\Gamma^{inh})^2} \quad , \quad (11)$$

where $\Gamma^{inh}$ is the inhomogeneous widening caused by crystal defects; $\Gamma_0^h = g^2[\hbar\omega(\Delta E_{exc} - \hbar\omega)]^{1/2}$ is homogeneous widening caused by optical phonons at T = 0; $\beta = 1+[(\Delta E_{exc} +\hbar\omega) / | \Delta E_{exc} - \hbar\omega |]^{1/2}$; $g$ is the constant of exciton-phonon interaction; $\Delta E_{exc}$ and $\hbar\omega$ are the exciton binding energy and phonon energy, respectively; and $n^*(T) = [\exp(\hbar\omega/kT) -1]^{-1}$ is a value which characterizes the filling of the phonon branch with temperature.

In accordance with [21], the homogeneous widening of the exciton absorption band in bulk InSe single crystals and the shift of the forbidden gap with the temperature growth occur with the participation of the homopolar half-layer $A^1_{1g}$ -phonon (ν = 110 cm$^{-1}$), and the homogeneous half-width $\Gamma_0^h$ for $n = 1$ exciton absorption line at T = 0 is $\Gamma_0^h$ = 1.3 meV [17]. In the case of InSe nanocrystals, when an exciton is scattered by a polar half-layer E-phonon (ν = 41 cm$^{-1}$), the value $\Gamma_0^h$ reaches 1.8 ± 0.2 meV in accordance with Eq. (11).

For comparison, Fig. 5c shows the PL spectrum of the bulk InSe single crystal containing an FE line of free exciton emission and a BE band corresponding to the set of lines for emission of bound excitons, the fitted G1 and G2 lines of bound exciton emission, and the L6 line of free exciton emission for InSe nanocrystals of radius $R^{QD}$ = 627 nm along with the total L6+G1+G2 band. The energy position of the L6 line maximum $E_{L6}$ = 1.3375 eV practically coincides with the emission maximum of free excitons in the bulk InSe crystal, $E_{n=1}$ = 1.3371 eV.

As can be seen from Fig. 5c), emission related to excitons bound at the boundary of InSe nanocrystals results in a spectrum which is qualitatively close to the emission spectrum of the InSe bulk crystal. Fig. 5c) also shows that the total intensity of radiative recombination in InSe nanocrystals, the concentration of which does not exceed 0.1% in the $In_2Se_3$ matrix, is comparable in magnitude with that in bulk InSe crystals.

It should be noted that the fitted half-width for the L6 line (see Table 2), $\Gamma_{L6} = 7$ meV, is four to five times higher than the calculated value of $\Gamma_0^h$ for the homogeneously widened ground state of bulk InSe crystal at T = 0. In crystals of radius < 150 nm, the fitted values $\Gamma_{Li}$ are already one order higher than $\Gamma_0^h$. At the same time, for bulk InSe crystals containing only a small number of lattice defects, the experimentally observed half-width of the emission line (Fig. 5c) and the absorption line [17] for free excitons at T = 5 K does not exceed 2–3 meV, which is only 1.5–2 times higher than $\Gamma_0^h$.

This widening of the emission line of free excitons is indicative that this exciton in InSe nanocrystals is subjected to an essential additional scattering, which results in non-homogeneous widening of the emission line. For L3–L5 lines, this non-homogeneous widening reaches $\Gamma_{Li}^{inh} = 14.9$ meV in accordance with Eq. (11) and provides the main contribution to the widening of the emission lines of free excitons at T = 0.

Note that the sharp increase in the contribution from non-homogeneous widening to the emission band of free excitons in InSe nanocrystals of radii less than 150 nm, as compared with bulk crystals, cannot be explained by scattering from defects available inside nanocrystals, since defects extend to the surface at these dimensions.

### 4.3.1. Photoluminescence of InSe QD in $In_2Se_3$ confinement at T = 5 K

Analytical calculations are presented in this and the following sections, where the radii of quantum dots are fixed values obtained in accordance with Eq. (7) when fitting the emission spectra of B and B1 bands by the set of emission lines of free and bound excitons.

It is assumed that when considering the secondary emission of light by a crystal (PL process) in the case of $R^{QD} \to r_{exc}$ of the bulk crystal, it is important to take into account the polariton model offered by Hopfield [45] and developed in [46-48] for the propagation of light in crystals, within the ranges of excitonic resonances.

In accordance with [45], the light wave propagating along the crystal within the range of excitonic resonances in the infinite ideal crystal is alternately either in the photonic or the excitonic state. In the general case, the plasma frequency of exciton-photon transition is as follows, according to [49]:

$$\omega_p = \sqrt{\omega_{exc} \cdot \omega_\Omega} \quad , \quad (12)$$

where $\omega_{exc}$ is the frequency for n=1 excitonic transition, and $\omega_\Omega$ is the frequency of crystalline vibration which takes part in exciton scattering. According to Eq. (12), the plasma frequency for the InSe crystal is $\omega_p = 2.05 \times 10^{14}$ s$^{-1}$; from this, the polariton free path in the excitonic state is $L_{e-p} = c/n_o\omega_p = 450$ nm.

From the other viewpoint, it was reported in [50] that light propagation in the vicinity of excitonic resonances in GaN essentially slows down, and the effective group velocity $v_{gr}$ drops to $2.1 \times 10^6$ m/s, which in the case of an InSe crystal gives the free path of excitonic polariton as $L_{e-p} = \hbar v_{gr}/2\Gamma_0^h = 506$ nm, where $\Gamma_0^h = 1.3$ meV [17].

It is known that in InSe crystals, so-called "thickness effects" (related to the decay of excitonic polaritons at T = 0, which leads to an increase in integrated intensity of the absorption excitonic band) begin to develop at a thickness of less than 40–50 mkm [17,51]. It therefore follows that, in the ideal infinite crystal, exciton decay takes place approximately once per 150 periods of a light wave in the excitonic state. Increases in the crystal temperature and number of defects, and a decrease in the geometrical sizes of crystals, result in a widening of the excitonic state.

Fig. 9 presents the data (for T = 5 K) from Table 2 (shown in circles), illustrating the dependence of $\Gamma(D)$, the half-width at the half-height for the emission excitonic line, on a reduction in the linear dimensions of InSe crystals from $L_{bulk} = 45$ mkm to $L = 2R^{QD} = 32.2$ nm. For convenience, in the following analysis the linear dimension of the crystals L is expressed in the units of exciton diameter $D = L/d_{exc}$, where $d_{exc} = 9.0$ nm is the exciton diameter in bulk InSe. As can be seen from Fig. 9, the data in Table 2 are well approximated at T = 0 by the analytical dependence of exciton line half-width on the crystal size, expressed in D units:

$$\Gamma_0(D) = \sqrt{[\Gamma_0^h]^2 + \left[\frac{\Lambda \cdot \Gamma_0^h}{\sqrt[3]{D}-1}\right]^2} \quad , \quad (13)$$

where $\Lambda$ is the analytical value of 23 for InSe, and $\Gamma_0(D)$ is dependent on D the half-width of the exciton line for absorption/emission in pure crystal at T = 0.

The analytical dependence expressed in Eq. (13) may be reduced to the well-known Davydov equation obtained for light propagation in the vicinity of exciton resonance (see [48]), and can be rewritten as

$$\Gamma_0(D) = \hbar\gamma^h(D) = \hbar\sqrt{[\gamma_0^h]^2 + \frac{2\omega_p^2}{4\varepsilon \cdot [\sqrt[3]{D}-1]^2}}. \quad (14)$$

Combining Eqs (13) and (14), the fitting parameter A in Eq. (13) can be expressed as

$$\Lambda = \hbar\omega_p \Big/ \sqrt{2}n_0\Gamma_0^h \quad . \quad (15)$$

The second component in Eqs. (13) and (14) represents a correction to the exciton half-width that is associated with a finite crystal size. In limiting cases, i.e. for $D \to \infty$, the value $\Gamma_0 \to \Gamma_0^h$, and for $D \to 1 = d_{exc}$, $\Gamma_0 \to \infty$.

In Fig. 9, the arrow indicates the free path corresponding to one period of the light wave in the excitonic state $L_{e-p} = 450$ nm, expressed in dimensionless units $D = L/d_{exc}$.

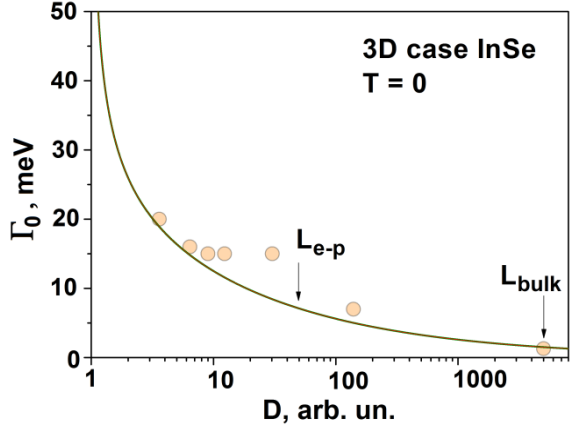

Fig. 9. Dependence of $\Gamma_0$ of the emission exciton line on D for pure InSe crystal at $T = 0$

In accordance with Eq. (13), when $\Gamma = 100$ meV the diameter of the InSe nanocrystal is 2.7% higher than that for exciton diameter in the bulk InSe crystal, and its lifetime $\tau = 3.29 \times 10^{-15}$ s should indicate recombination of a non-bound electron-hole pair.

### 4.3.2. Photoluminescence of the InSe QD ensemble in $In_2Se_3$ confinement at T = 5 K

Earlier electron-microscopic and optical investigations of GaSe crystals intercalated in a $RbNO_3$ melt [14] showed that the concentration of $RbNO_3$ nanocrystalline inclusions can reach 10%, while the segmented polycrystal GaSe matrix retains its optical properties.

When comparing the light absorption coefficients at T = 5 K in the range < 1.45 eV (Fig. 2a indicates the range of light absorption in InSe nanocrystals) and for absorption near the band-to-band transitions in $In_2Se_3$ matrix, it can be assumed (by analogy with [14]) that the concentration of InSe nanocrystals in $In_2Se_3$ matrix is less than 0.1%.

At the same time, the juxtaposition of the PL spectra shown in Fig. 5c) for the bulk InSe crystal and the calculated L6, G1 and G2 emission lines obtained for InSe nanocrystals under practically identical experimental conditions confirms the high efficiency of radiative recombination for InSe nanocrystals confined in the $In_2Se_3$ matrix.

In view of the above, the resultant half-width of the B band for the ***different intensity*** $l = 6$ lines of free exciton emission and $m = 2$ emission lines of bound excitons at T = 0 is equal to

$$\Gamma_0^B = \sqrt{\sum_{i=1}^{l} M_i \Gamma_{Li}^2(0) + \sum_{j=1}^{m} N_j \Gamma_{Gj}^2}, \quad (16)$$

where $M_i$, $N_j$ and $\Gamma_{Li}(0)$, $\Gamma_{Gj}$ are weighting coefficients defining the intensities and half-widths of fitted emission lines. As can be seen from Fig. 6 and Table 2, the main contribution at T = 5 K (up to 95%) to the formation of B and B1 bands is made by the L2–L5 and G1 fitted emission lines.

Therefore, neglecting the contributions of the L1, L6 and G2 lines, according to Eq. (16) and given in Table 2 values of half-widths for L2–L5 and G1 fitting lines (when the weighting coefficients $M_2$–$M_5$ and $N_1$ are equal to unity), the following calculated values of half-widths can be obtained for the B and B1 bands: $\Gamma_B = 44.6$ meV and $\Gamma_{B1} = 42.7$ meV. These coincide with good accuracy with the experimental data at T = 5 K: $\Gamma_B = 45.0$ meV and $\Gamma_{B1} = 40.5$ meV, as given in Table 1.

Based on the corrections to the half-widths of the L2–L5 and G1 lines, obtained according to Eqs. (13) and (14), half-width values of $\Gamma_B = 43.1$ meV and $\Gamma_{B1} = 40.9$ meV can be deduced. Taking into account that the G2 line also gives a non-zero contribution (approximately 15% relative to G1-line) to the emission intensity of the B band, there is practically full correspondence between the fitted L1–L6, G1and G2 lines and the emission spectra of the B and B1 bands at T = 5 K.

### 4.3.3. Photoluminescence of the InSe QD ensemble in $In_2Se_3$ confinement at T > 5 K

It is notable that essential growth of free exciton emission line half-width occurs when the sizes of the InSe crystal become smaller than the free path of the polariton in the exciton state $L_{e-p}$, for the case of the ideal infinite crystal.

To investigate the influence of temperature on widening the exciton lines of absorption/emission in nanocrystals, we separate the half-width using two components, homogeneous and non-homogeneous, as for the case of bulk crystals. As an initial homogeneous component, one can take the homogeneous half-width $\Gamma_0^h = 1.3$ meV of the absorption exciton line in the bulk InSe crystal. Then, based on the half-width values given in Table 2 for the L3–L5 lines, we can obtain the value of the non-homogeneous component as $\Gamma_{Li}^{inh} = 14.9$ meV.

Fig. 10 shows the experimental dependences of the half-width $\Gamma$ on temperature for the B and B1 bands (shown with squares and circles), taken from Table 1. The same figure shows curve 1, plotted in accordance with Eqs. (11) and (16) and parameters $M_{1,6} = 0$; $N_2 = 0$; $M_{2,3,4,5} = 1$; $N_1 = 1$; $\Gamma_{L1,L2,L3,L4,L5}^{inh} = 14.9$ meV; $\Gamma_{G1}^{inh} = 35.0$ meV. It is obvious that curve 1 only qualitatively describes the widening of band B with increasing temperature.

An additional component providing a contribution to the widening of the B band with temperature can be explained by taking into account the fact that the kinetic energy of excitons $\hbar^2 k^2/2M$ (where $M = m_e + m_h$), being closed inside a sphere of radius less than the free path of polaritons in the exciton state, is increased; this results in an increase in the frequency of exciton scattering on the walls of a sphere with radius $R^{QD}$.

In our case, a 3D exciton gas, the exciton kinetic energy (similarly to that in an ideal gas) is proportional to $(3/2)kT$. Then, Eq. (16), based on Eqs (11) and (13), can be transformed to a general form:

$$\Gamma^{B,B1}(T,R) = \sqrt{\sum_{i=1}^{l} M_i\{[\Gamma_{L_i}^h(T,R)]^2 + [^3/_2\,\Theta_i(R)\cdot kT]^2 + [\Gamma_{L_i}^{inh}(R)]^2\} + \sum_{i=1}^{l} M_i\{\sum_{j=1}^{m} N_j \Gamma_{G_j}^2(R)\}}\,, \quad (17)$$

where $\Theta_i = f(R^{QD})$ is the fitting parameter, which increases with a decrease in the radius of nanocrystal.

Taking into account the additional term in Eq. (17) (as an approximation, $\Theta_i = const = 1.3$) for the B and B1 bands, we obtain curves 2 and 3 (see Fig. 10), which describe the experimental dependences $\Gamma_{B,B1}(T)$ relatively well up to $T < 80$ K. Note that curve 3 for the B1 band was obtained using Eq. (17) at the same parameters L2–L5 that were used for B band fitting, although the half-width was taken as $\Gamma_{G1} = 30.0$ meV.

In summary, the parameter $\Theta_i = f(R^{QD})$ was fitted to the experimental dependence $\Gamma_{B,B1}(T)$ and to the analytical curve 2 (Fig. 10) obtained in accordance with Eq. (17) for $\Theta_i = const = 1.2$. Using these four lines, L2–L5, which provide the main contribution to the half-width, the integrated intensity of B and B1 bands and the radii shown in Table 2, we obtain self-consistent values for the parameter $\Theta_i$, that can be described by the simple analytic function

$$\Theta_i(R) = \vartheta\,\frac{r_{exc}}{R}\,, \quad (18)$$

where $\vartheta = \omega_{cv}/\omega_p$, and $\omega_{cv}$ is the frequency of fundamental electron transition between the conductive and valence bands.

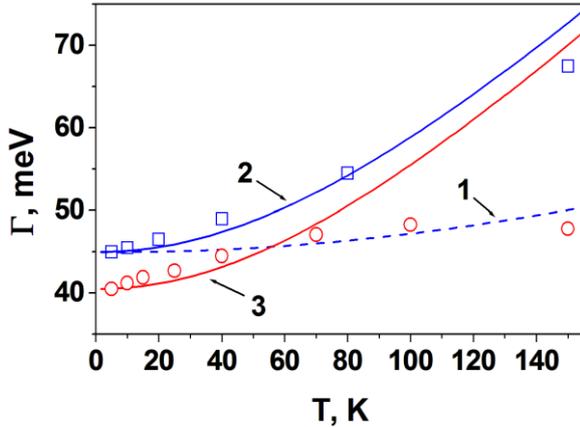

Fig. 10. Dependences of the half-widths $\Gamma$ for the B band (squares) and B1 band (circles) on temperature. Curve 1 corresponds to Eq. (11) and curves 2 and 3 to Eq. (17)

Fig. 11 shows the obtained values for four fitted radii and curve 1, plotted in accordance with Eq. (18). Curve 1 has finite values: for $R \to \infty$, the parameter $\Theta \to 0$, and for $R \to r_{exc}$, $\Theta \to \vartheta$. It can be seen from Eqs. (13) and (14) that the highest contribution to the widening of the emission line of a free exciton with an increase in temperature occurs at $R^{QD} < 55$ nm. In the case where $R^{QD} = 5$ nm (for $r_{exc} = 4.5$ nm), the contribution of kinetic energy to the widening of the exciton band is 0.47 meV for T = 5 K and 28.2 meV for T = 300 K.

It should be also noted that for $R^{QD} = 5$ nm, the value of non-homogeneous widening is equal to 61.5 meV, in accordance with Eqs. (13) and (14). It therefore follows that the main contribution to the widening of the exciton band of InSe nanocrystals (with diameter equal to 2–5 exciton diameters) up to room temperature is provided by non-homogeneous widening, which depends on the crystal radius.

The curve 3 in Fig. 10 illustrates the temperature behavior of the half-width $\Gamma$ for the emission B band consisting of the ensemble of L2–L5 emission lines of free excitons and the G1 and G2 emission lines of bound excitons. Curve 3 was obtained by fitting the emission lines in accordance with Eqs. (13), (17) and (18) with weighting coefficients $M_i$, $N_i$ for the fixed values $R_i$, obtained using Eq. (7). Curve 2 is identical to curve 3, and was plotted according to Eq. (17).

In the general case for $\Gamma(R,T)$ dependence, it follows from Eq. (17) that one should take into account that each emission L line of free excitons with radius $R$ corresponds to the emission $G_j$ line of bound excitons with the same weighting coefficient $M_j$.

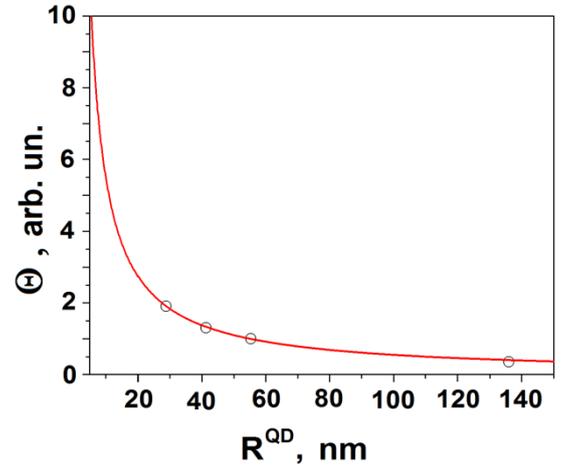

Fig. 11. Dependence of the $\Theta$-parameter on the radius of InSe nanocrystal

It is noteworthy that the emission lines of bound excitons observed in the G1 and G2 fitting spectra (see Table 2) have different half-widths and integrated intensities, and are shifted in energy relative to each other. As shown by electron-microscopic investigations, it can be assumed that the existence of two lines indicates their different geometrical shapes. The G1 line corresponds to the emission of excitons bound at the spherical boundaries, while G2 corresponds to those at the ellipsoidal ones. It also should be noted that the concentration of spherical InSe nanocrystals in the $In_2Se_3$ matrix is, on the whole, one order higher than that of the ellipsoidal ones.

We turn our attention to the behavior of the half-widths of the B and B1 emission bands with increasing temperature, and in particular the lag in experimental values at $T > 80$ K from the curve plotted in accordance with Eq. (17).

It seems reasonable to assume that for increasing temperature, this lag is related to the gradual overlap between the upper conduction minibands of InSe nanocrystals and the conduction band of the matrix $In_2Se_3$. This effect is more clearly pronounced for the B1 band, since the integrated intensity of the G1 emission line in the B1 band is essentially less than that of the B band, as can be seen from Fig. 5 and Table 1.

Moreover, as shown in Fig. 9, the experimental value of the B1 band half-width begins to decrease at T > 100 K. This fact corresponds with the appearance shown in Fig. 2c) and the subsequent increase in intensity of A-band emission at T > 80 K when simultaneously decreasing the B-band intensity.

### 4.3.4. Emission of free excitons in the bulk $In_2Se_3$ crystal

Based on the above discussion of the bulk $In_2Se_3$ crystal, we can determine the energy of an optical phonon $\hbar\omega = 6.0$ meV taking part in reducing the forbidden gap width with increasing temperature, and the parameter for exciton (electron)/phonon interaction $g = 0.65$ ($g_c = g_v = 0.65$). By analogy with the previous calculations, this allows us to ascertain the value of homogeneous widening for free excitons in the $In_2Se_3$ crystal at $T = 0$, which is equal to $\Gamma_0^h = 2.93$ meV, with coefficient $\beta = 2.58$, and to obtain the analytical dependence for homogeneous widening of the free exciton lines in the bulk $In_2Se_3$ crystal

$$\Gamma^h(T)[meV] = 2.93 \cdot [1 + \frac{2.58}{\exp(^{93}/_T)-1}] \quad . \quad (19)$$

Therefore, using the PL spectrum of the $In_2Se_3$ crystal at $T = 200$ K (see Fig. 2c), we can determine the experimental half-width of the free exciton emission line at this temperature as $\Gamma = 21$ meV, and in accordance with Eqs (11) and (19) we can determine the homogeneous $\Gamma^h(200\ K) = 17.3$ meV and non-homogeneous $\Gamma^{inh} = 11.9$ meV components respectively, for a free exciton emission line. A large value of $\Gamma^{inh}$ is evidence of a significant number of crystal structure defects in the bulk $In_2Se_3$ crystal.

Shown in the same figure is the emission line, with a maximum at 1.0 μm and half-width $\Gamma = 77 \pm 2$ meV, located 65 meV lower than the emission line of free excitons. The temperature shift of this band is different from $E_g$, and this indicates that nature of this band is related to the recombination of electrons from shallow donor levels (located 20–25 meV below the bottom of the CB) to the valence band, with the participation of two to three optical A phonons.

### 5. Conclusions

SEM, EDX, and optical investigations of Bridgeman-grown $In_2Se_3$ crystals conducted here show that bulk $In_2Se_3$ crystals as a whole have a homogeneous concentration of In and Se atoms, corresponding with $In_2Se_3$ stoichiometry.

At the same time, these crystals contain dislocations of 100–200 nm in radius; in the vicinity of these dislocations, the interspersion of foreign sub-micron "pocket" phases of 400–600 nm in size of β-Se, InSe and $In_6Se_7$ crystals was observed.

In addition, $In_2Se_3$ contains much smaller, nanosized dislocation veins of 15–30 nm in radius, with spherical and ellipsoidal inclusions located on them with radius r = 30 nm or main axes a = b = 30 nm and c = 150 nm, respectively.

Using temperature investigations of emission and the fundamental edge absorption spectra of $In_2Se_3$ crystals, the following were found:
- The band gap width, $E_g = 1.550 \pm 0.001$ eV, at T = 0;
- The exciton binding energy, $\Delta E_{exc} = 14 \pm 1$ meV;
- The dielectric permeability of the crystal in region of fundamental transition, $\varepsilon = 9.9 \pm 0.1$;
- The exciton-phonon coupling constant, $g = 0.65$ ($g_c = g_v = 0.65$);
- The effective masses of carriers, $m^*_c = 0.12\ m_0$ and $m^*_v = 0.6\ m_0$;
- The frequency ($\nu_1 = 48 \pm 3$ cm$^{-1}$) of the half-layer homopolar A-phonon which takes part in (electron)exciton-phonon interaction in $In_2Se_3$ crystals.

It is shown that the PL spectra observed at low temperatures in $In_2Se_3$ crystal bands, referred to here as B and B1 emission bands, may be approximated by the set of free exciton L1–L6 emission lines in InSe nanocrystals of 16 to 650 nm in radius, and the G1 and G2 emission lines of excitons bounded at boundaries between InSe nanocrystals and the $In_2Se_3$ matrix.

The G1 line corresponds to the emission of excitons bound at the spherical boundaries, while G2 refers to those at the ellipsoidal ones. It should also be noted that the concentration of spherical InSe nanocrystals in the $In_2Se_3$ matrix, on the whole, is one order higher than that of the ellipsoidal ones.

A scheme is offered of the energetic displacement of the valence and conductive minibands of InSe nanocrystals of different radii according to the conductive and valence bands of the $In_2Se_3$ matrix and bulk InSe crystals, and the electron recombination channel at T = 0 and T = 300 K.

It was found that exciton-phonon interaction in InSe nanocrystals can be attributed to participation by a half-layer polar E-phonon ($\nu = 40 \pm 1$ cm$^{-1}$). This is the own vibration of InSe crystal and is in resonance with the half-layer homopolar A phonon ($\nu_1 = 48$ cm$^{-1}$) of the $In_2Se_3$ matrix.

For both InSe nanocrystals and bulk InSe crystals, the decrease in the band gap width with an increase in temperature is not as rapid as in the $In_2Se_3$ crystal matrix. Thus, in PL spectra at T > 80 K, a switch in the recombination channel takes place and an A band of free exciton emission appears on the background of the B and B1 bands of the $In_2Se_3$ matrix. For PL spectra at T > 200 K, only the emission of the free exciton line of the $In_2Se_3$ matrix and wide band is observed. The nature of this band is related to the recombination of electrons from shallow donor levels (located at 20–25 meV below the bottom of the $In_2Se_3$ CB) to the valence band with the participation of two to three optical A phonons.

Using the framework of ideal exciton gas and classical exciton-phonon interaction models, simple analytical dependencies were found that connect an increase in exciton half-width with:

- A decrease in exciton free path when the size of the bulk InSe crystal is reduced to the size of its exciton diameter;

- An increase in free exciton kinetic energy proportional to $kT$, when: i) increasing the InSe 3D nanocrystal size at fixed temperature; and ii) increasing the temperature for a fixed nanocrystal size.

Contrary to bulk single crystals, where the main contribution to the increase in exciton line half-width is attributed to exciton-phonon interaction, in InSe nanocrystals with size between two and five exciton diameters, the main contribution to the increase in exciton line half-width up to room temperature is attributed to exciton kinetic energy and inhomogeneous broadening, and is strongly dependent on nanocrystal radius.

On the basis of the investigations conducted here and calculations of ensembles with different radii of InSe nanoncrystals confined in the $In_2Se_3$ matrix, the general dependence was obtained of the emission band half-width (made up of free and bound exciton emission lines in the InSe nanoncrystals ensemble) on temperature and size.

**Acknowledgment**

**References**

[1] L. Britnell, R. V. Gorbachev, R. Jalil, B. D. Belle, F. Schedin, A. Mishchenko, T. Georgiou, M. I. Katsnelson, L. Eaves, S. V. Morozov, N. M. R. Peres, J. Leist, A. K. Geim, K. S. Novoselov, L. A. Ponomarenko, *Science* **2012**, *335*, 947.

[2] Sukrit Sucharitakul, Nicholas J. Goble, U. Rajesh Kumar, Raman Sankar, Zachary A. Bogorad, Fang-Cheng Chou, Yit-Tsong Chen, and Xuan P.A. Gao, Nano Letters, 15 (6), pp. 3815-3819 (2015)

[3] P. Hu, Z. Wen, L. Wang, P. Tan, and K. Xiao, ACS Nano **6,** 5988 (2012).

[4] Z. Yang, W. Lie, C.H. Mak, S. Lin, H. Lin, X. Yang, F. Yan, S.P. Lay, J. Hao, ACS Nano 2017, 11,4225-4236.

[5] S.R. Tamalampudi, Y.Y. Lu,U.R. Kumar, R. Sankar, C.D. Liao, B.K. Moorthy, C.H. Cheng, F.C. Chou, Y.T. Chen, Nano Letter, 2014, 14, 2800-2806.

[6] S. Lei, L. Ge, S. Najmaei, A. George, R. Cappera, J. Lou, M. Chowalla, H. Yamaguchi, G. Gupta, R. Vajtai, A.D. Mohite, P.M. Ajayan, ASC Nano, 2014, 8, 1263-1272.

[7] J.O. Island, S.I. Blanter, M. Buscema, H.S.J. van der Zant, and A. Castellanos-Gomes, *Nano Lett.*, **2015**, *15* (12), pp. 7853–7858, **DOI:** 10.1021/acs.nanolett.5b02523.

[8] W. Feng, J.B. Wu, X. Li, W. Zheng, X. Zhou, K. Xiao, W. Cao, B. Yang, J.C. Idrobo, L. Basile, W. Tian, P. Tan, P. Hu, J. Matter. Chem. C 2015, 3, 7022-7028.

[9] Nguen T. Hung, Ahmad R.T. Nugraha, and Riichiro Saito, arXiv:1705.06688v3, 24 Jul, 2017

[10] S. Bertolazzi, D. Krasnozhon, A. Kis, *ACS Nano* **2013**, *7*, 3246.

[11] M.K. Song, S.J. Park, F.M. Alamgir, J.P. Cho, and M.L. Liu, Material Science and Engineering R; Reports R72, 2004 (2011).

[12] Yuriy Zhirko, Volodymyr Trachevsky, and Zakhar Kovalyuk, in book: "Hydrogen storage", by Editor Jianjun Liu, ISBN 978-953-51-0731-6, InTech, 136 pages, 2012, Chapter 2, p.211-242.

[13] Yu.I. Zhirko, V.M. Grekhov, N.A. Srubenko, Z.D. Kovalyuk, T.M. Feshak, in book "Advanced Materials for Renewable Hydrogen Production, Storage and Utilization", by Editor Jianjun Liu, ISBN 978-953-51-4273-7, InTech, 136 pages, 2015, Chapter 2, p. 11-50.

[14] Yuriy Zhirko, Vasiliy Grekhov, Zakhar Kovalyuk, Victor Netyaga, International Journal of Engineering Research & Science (IJOER), Vol-3, Issue-8, August- 2017, pp. 12 – 19, ISSN: 2395-6992, DOI: 10.25125/engineering-journal.

[15] A.V. Saslonkin, Z.D. Kovalyuk,I.V. Mintjanskii, Neorganocheskie Materialy, 43, 1415, 2007.

[16] M.D. Yang, C.H. Hu, S.C. Tong, J.L. Shen S.M. Lan, C.H. Wu, and T.G. Lin, Journal of Nanomaterials, Volume 2011, Article ID 976262, 5 pages, doi: 10.1155/2011/976262.

[17] Yu.I. Zhirko, Phys. St. Sol. (b), 213, 93, 1999.

[18] K. Osamura, Y. Murakami, Y. Tomiie, J. Phys. Soc. Japan, 21, 1848, 1966.

[19] Balkanski, C. Julien, A. Chevy, K. Kambas, Sol. St. Commun. , V. 59 , No 7, P. 423-427 (1986).

[20] S. Nagai, Sh. Saito, K. Nakao, J. Phys. Soc. Japan, 64, 1622

[21] J. Camassel, P. Merle, H. Mathieu, A. Chevy, Phys. Rev.:B, 17, 4718, 1978.

[22] Ph. Schmid, J.P. Voitchovsky, Phys. St. Sol. (b), 65, 249, 1974.

[23] J.M. Besson, Nuovo Chemento, B38, 478, 1977.

[24] D. Olguin, A. Cantarero, C. Ulrich, K. Suassen, Phys. St. Sol. (b), 235, 456, 2003.

[25] N. Kuroda, Y. Nishina, J. Phusica, 105B, 30, 1981.

[26] L.N. Alieva, R.Kh. Nani, E. Yu. Salaev, G.L. Belenkii, I.I. Reshina, V.Ja. Shteinshrajber, Phys. St. sol. (b), 1977, 82, No2, p.705-709.

[27] In book "Ternary Compounds, Organic Semiconductors" by O. Madelung, U. Rössler, M. Schulz, (ISBN 978-3-540-66781-0), Springer-Verlag, 2000.

[28] C. Julien, A. Chevy, D. Siapkas, Phys St. Sol. (a), 118, 553 (1990).

[29] T. Ikari, S. Shigetomi, K. Hashimoto, Phys. St. sol. (b), 1982, v.111, No2, p. 477-481.

[30] C. Carlone, S. Jandl, .H. Shanks, Phys. St. sol. (b), 1981, v.103, No1, p. 123-130.

[31] Yu.P. Gnatenko, Yu.I. Zhirko, Z.D. Kovalyuk, V.M. Kaminskii, Fis. Tverd. Tela, v.18, No 12, p. 3591-3594, 1986.

[32] J.J. Forney, R. Machke, E. Moozer, Nuovo Cimento, 1977, v. 38B, No2, p. 418-422.

[33] A. Kazuya, I. Sasaki, S. Hashimoto, Y. Nishina, H. Iwasaki, Sol. St. Commun., 1985, v.55, No1, p. 63-66.

[34] Y. Sasaki, Y. Nishina, Physica, 1981, v.BC105, No1-3, p. 45-49.

[35] E. Kress-Rogers, G.F. Hopper, R.G. Nikolas, W. Hayes, J.C. Portal, A. Chevy, J. Phys. C: Solid State, 1983, v.16, p. 4285-4295.

[36] N. Kuroda, Y. Nishina, Sol. St. Comm., 1980, 34, p. 481-484.

[37] Y. Wang, A. Suna, W. Mahler, and R. Kasowski, Journal of Chemical Physics, 87(12), 7315-7322, 1987.

[38] Yu.I. Zhirko, in book: "Feeling in Future: Advances in Science and Technologies for Energy Generation, Transmission and Storage" by Editor A. Mendez-Vilas, ISBN-13: 978-1-61233-558-2, BrownWalker Press, December, 2012, p.558-562.

[39] D.A. Bandurin, A.V. Tyurnina, G.L. Yu, A. Mishenko, V. Zolyomi, S.V. Morozov, R. Krishna-Kumar, R.V. Gorbachev, Z.R. Kudrinskyi, S. Pezzini, Z.D. Kovalyuk, U. Zeitler, K.S. Novoselov, A. Patane, L. Eaves, I.V. Grigorieva, V.I. Fal'ko, A.K. Geim, Y. Cao, Nature Nanotechnology 12, 223-227 (2016).

[40] S. J. Magorrian, V. Zolyomi, and V. I. Fal'ko, arXiv:1611.00262v2 [cond-mat.mes-hall] 2 Dec 2016.

[41] Garry W. Mudd, Simon A. Svatek, Tianhang Ren, Amalia Patanè, Oleg Makarovsky, Laurence Eaves, Peter H. Beton, Zakhar D. Kovalyuk, George V. Lashkarev, Zakhar R. Kudrynskyi, and Alexandr I. Dmitriev, Adv. Mater. 2013, 25, 5714–5718, DOI: 10.1002/adma.201302616.

[42] Yu.I. Zhirko, 14-th European Conference of Solid State Chemistry, "ECSSC14", July, 7-10, 2013, Bordeaux, France, p.139.

[43] R.J. Elliott, Phys. Rev. 108, 1384, 1957.

[44] Y. Toyozawa, Progr. Theor. Physic, 1958, v.20, No1, p. 53-81.

[45]. J.J. Hopfield, Phys. Rev. 112, 1955, 1958.

[46] J.J. Hopfield & D.G.Thomas, Phys. Rev., **132**, 563, 1963.

[47] S.I. Pekar, Zh. Exper. Theor. Phys., 33, 1022, 1957.

[48] A.S. Davydov and A.A. Serikov, Phys. St. Sol. (b), 14, 933, 1974.

[49] Yu. I. Zhirko, unpublished.

[50] T.V. Shubina, Thesis of Dr. Sci. dissertation, A. Ioffe Physic-technical institute, 2008, p.43.

[51] L.N. Kurbatov, A.I. Dirochka, V.A. Sosnin, Fiz. Tekh. Polupr. (in Russian), 1979, v. 13, No1 p.83-88.

[52] Yuriy Zhirko, Nikolay Skubenko, Volodymyr Dubinko, Zakhar Kovalyuk, Oleg Sydor, Journal of Materials Science and Engineering B, 2013, No3, P.162-174.